# Near-field optomechanical transduction enhanced by Raman gain

RYOKO SAKUMA*, MOTOKI ASANO, HIROSHI YAMAGUCHI, AND HAJIME OKAMOTO

*NTT Basic Research Laboratories, NTT Corporation, 3-1 Morinosato Wakamiya, Atsugi-shi, Kanagawa 243-0198, Japan*
**Ryoko.sakuma@ntt.com*

**Abstract:** Raman-gain-enhanced near-field optomechanical transduction between a movable optical cavity and SiN-membrane resonator is demonstrated. The Raman gain compensates for the intrinsic loss of the cavity and amplifies the optomechanical transduction, through which the membrane vibration is sensed using a high-Q whispering-gallery-mode optical cavity evanescently. The optical Q of the cavity resonance is improved with respect to the optical pump power, which results in an increase in the optomechanically transduced vibration signals of the mechanical resonator. Our near-field optomechanical coupling approach with optical gain realizes highly sensitive displacement measurement in nano- and micro-mechanical resonators consisting of arbitrary materials and structures.

## 1. Introduction

Cavity optomechanical systems have been developed to measure and control small mechanical motions in nanomechanical systems. These capabilities are often achieved with high-Q optical microresonators including Fabry-Pérot cavities, photonic crystal cavities, and whispering-gallery-mode (WGM) resonators, wherein optomechanical coupling transduces mechanical motion to a frequency shift of the cavity mode [1–4]. The resulting dispersive frequency shift enables super-sensitive readout of mechanical motions owing to the ultranarrow cavity resonance [5]. Recently, the position-measurement precision has achieved a sensitivity of the standard quantum limit [3,6]. These super-sensitive displacement measurements in high-Q optical microresonators open the way to various types of sensing applications, including force [7,8], acceleration [9], torque [10], and gravitational wave detectors [11].

The optomechanical transduction increases with the intracavity light power and the inverse of the cavity linewidth (i.e., optical Q factor) [5]. Thus, an active optical process in which optical gain compensates for the losses has great potential to increase the sensitivity because the gain increases the optical Q, leading to improved transduced signals. The optical gain approach has been demonstrated in various types of cavity sensing (e.g., small particle detection [12]) by using rare-earth doping [13,14], nonlinear optical mixing [15], and Brillouin and Raman scattering [14,16]. To introduce the optical gain approach into optomechanical systems incorporating unified optomechanical resonators, the resonators must be fabricated by including a nonlinear or gain medium. However, this restriction significantly limits the candidates for resonator materials in terms of mechanical properties.

Here, we report the first demonstration of improvement in transduced optomechanical signals in nanomechanical motions using near-field optomechanical coupling with Raman gain. The near-field optomechanical configuration allows us to individually fabricate the optical and mechanical resonators [17]. The optical gain properties and mechanical vibration characteristics can be independently optimized because the resonators can be fabricated in any materials and structures. Various types of mechanical motions can be sensed, including motions of membranes [18], nanowires [17], double- and single-clamped photonic crystal beams [19], and vibrating molecules [20]. We performed proof-of-principle studies with a microsphere cavity with a high optical quality factor (Q = $4.2\times10^{7}$), and experimentally confirmed the improvement in the optomechanical transduction by increasing the optical gain. The Raman scattering process in a silica microsphere was used for the optical gain because this process has a wide excitation bandwidth that includes the telecommunications band. Therefore, the optomechanical transduction can be enhanced by using fiber-based telecommunications technology, which could be applicable for variety of optomechanical sensing scenarios.

## 2. Setup

### *2.1 Principle of Raman gain-enhanced optomechanical transduction*

A schematic representation of the near-field optomechanical coupling between a silicon nitride (SiN) mechanical membrane resonator and silica WGM microsphere cavity via Raman gain is schematically shown in Fig. 1(a). The probe and pump lasers are evanescently coupled to the resonator by a waveguide fabricated with a tapered fiber. The pump laser induces stimulated Raman amplification—energy is transferred from the pump laser to the Stokes photons. The probe and pump lasers are set to the appropriate frequencies that satisfy $\omega_{probe} = \omega_S = \omega_{pump} - \omega_v$, where $\omega_S$ and $\omega_v$ denote the Stokes frequency and phonon vibration frequency of the cavity medium.

The optical cavity and membrane resonator are independently positioned; it is not in an integrated system. This configuration allows free access to the mechanical vibration mode through the optical cavity. By making an overlap between the optical evanescent field of WGMs and the membrane resonator, the mechanical vibration of the resonator is transduced into the frequency shift of the cavity resonance owing to the near-field optomechanical coupling. The transduced signal is proportional to the change in the probe transmission, as described in Fig. 1(b), and thus the maximum optomechanical transduction is achieved when the probe-laser detuning corresponds to the half width at half maximum (HWHM) of the cavity resonance [21]. Assuming that the probe detuning is locked at the HWHM of the probe resonance, we can quantify the optomechanical transduction for the square of the mechanical displacement using transduction coefficient $\eta$. The transduction coefficient, which is described with intracavity probe power $P_{\mathrm{cav}}$[22], can be simplified to $\eta = P_{\mathrm{cav}}^2 Q^2$.

In a cavity optomechanical system, the total loss in the cavity with Raman gain can be described as a summation of effective intrinsic loss ($\kappa_{\mathrm{eff}}$) and coupling loss ($\kappa_{\mathrm{ex}}$) between the cavity and waveguide: $\kappa = \kappa_{\mathrm{eff}} + \kappa_{\mathrm{ex}}$. The amplification rate by the Raman gain is denoted by $\xi$, and the effective intrinsic loss can be given by $\kappa_{\mathrm{eff}} = \kappa_0 - \xi$, where $\kappa_0$ denotes the intrinsic loss without Raman gain [14]. Increasing the pump power compensates for the intrinsic energy loss. In the single optical mode, the coupling state where $\kappa_{\mathrm{eff}} = \kappa_{\mathrm{ex}}$ is referred to as critical coupling, in which the light transmission $T$ drops to zero and the input power is completely dissipated within the cavity. The situations $\kappa_{\mathrm{eff}} > \kappa_{\mathrm{ex}}$ and $\kappa_{\mathrm{eff}} < \kappa_{\mathrm{ex}}$ are called under- and over-coupling and refer to the states where the intrinsic or coupling loss dominates the total losses, resulting in a shallower transmission depth [5]. The coupling states gradually change from under- to over-coupling as the tapered fiber approaches the cavity, resulting in a larger $\kappa_{\mathrm{ex}}$.

Fig. 2(a) and (b) show the change in the calculated intracavity probe power $P_{cav}$, optical Q, and transduction coefficient $\eta$ at the under ($\kappa_{\mathrm{ex}}/\kappa_0 = 0.5$) and critical ($\kappa_{\mathrm{ex}}/\kappa_0 = 1.0$) coupling states with increasing Raman amplification rate $\xi$. All values are normalized with respect to that at the critical coupling state without Raman gain. The overall transduction coefficient is larger at the under-coupling state because of the sharply increasing optical Q. A relatively small coupling loss at the under-coupling state causes a larger increment in the Q factor, which is described as $Q = \omega_o/(\kappa_0 + \kappa_{ex} - \xi)$, where $\omega_o$ is the cavity resonance frequency. The intracavity probe power is maximized at the critical coupling state, whereas the linewidth narrows with increasing $\xi$. The optomechanical transduction efficiency can be maximized at the optimized coupling and optical gain in the slightly over-coupled regime ($\kappa_{\mathrm{eff}} < \kappa_{\mathrm{ex}}$) where the product $\eta = P_{\text{cav}}^2 Q^2$ becomes largest. The condition is experimentally achieved by balancing the degree of the cavity-waveguide coupling states and optical gain $\xi$ provided by the pump power.

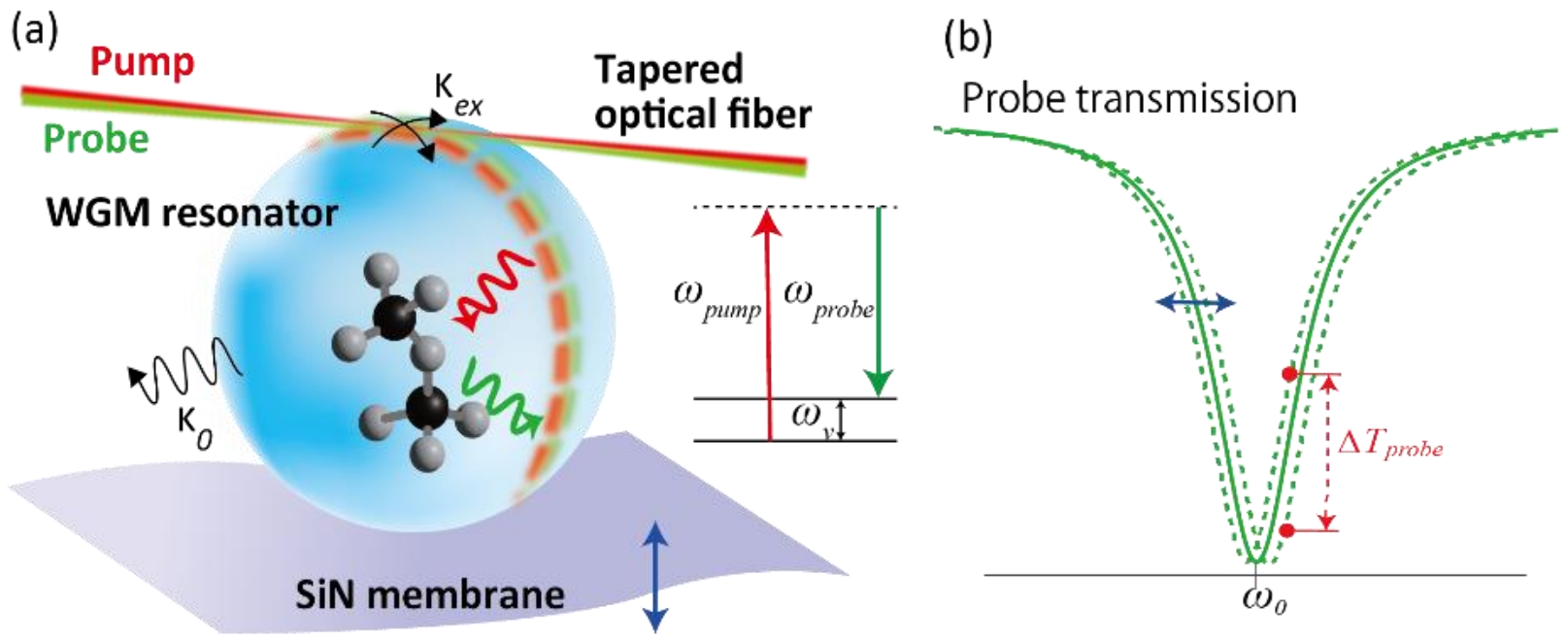

Fig. 1. (a) Schematic representation of Raman-gain enhanced near-field optomechanical transduction between an optical cavity and thin membrane resonator. The energy diagram shows the Raman process. (b) Schematic of a change in the probe transmission owing to membrane vibrations.

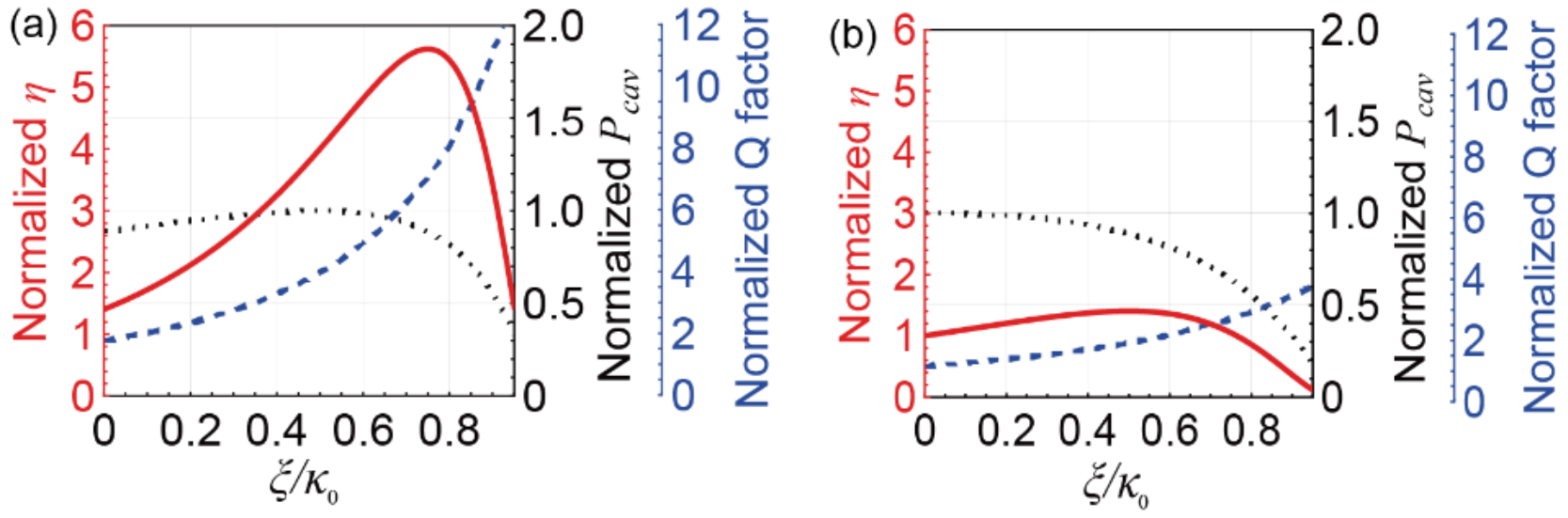

Fig. 2. Change in the calculated intracavity probe power $P_{cav}$, optical Q, and transduction efficiency $\eta$ at (a) under-coupling ($\kappa_{\text{ex}}/\kappa_0 = 0.5$) and (b) critical coupling states ($\kappa_{\text{ex}}/\kappa_0 = 1.0$) with increasing Raman amplification rate $\xi$. All values are normalized with respect to that at the critical coupling state with $\xi = 0$.

### 2.2 Experimental setup

A schematic illustration of the experimental setup is shown in Fig. 3. A silica microsphere cavity with a diameter of ~30 μm was prepared at the end of a silica fiber by the high-voltage electric discharge technique [23]. A tapered silica fiber with a diameter of ~2 μm was contacted to the cavity, and ~1570 and ~1475-nm lasers were used as the probe and pump light. Silica glass has a Raman response with the highest gain at 13.2 THz away from the pump frequency [24]. The spectrum range is broad; more than half of the maximum spectrum intensity can be obtained at 10–15 THz away from the pump frequency. The corresponding pump-wavelength range is 80–115 nm away from the probe wavelength for the ~1570 nm laser.

The pump power was set between 1.0 and 2.5 mW to adjust the optical Raman gain. The probe laser power was kept at 0.5 μW, which is small enough to avoid thermal broadening of the resonance dip. The pump frequency was tuned to maximize the pump-probe spatial overlap by monitoring the Raman response of the probe resonance. A large input power often changes the polarization state of a tapered fiber owing to the change in the refractive index generated by the thermal effect or nonlinear optical effect; therefore, polarization was adjusted by a fiber polarization controller (FPC) to maximize the transmission in every measurement.

The tapered fiber, silica microsphere, and SiN membrane resonator were placed in a chamber filled with nitrogen gas [see Fig. 3]. The xyz-position of the cavity relative to the tapered fiber can be precisely controlled by a piezo-positioner (attocube, ANC300) attached to the resonator, and it was determined so as to minimize the linewidth of the probe-resonance dip to ensure the coupling state is at the under-coupling condition. A membrane resonator (50×50 μm) with a thickness of 30 nm was placed on a PZT (lead zirconate titanate) transducer, and a random noise of 200 mV amplitude was applied to the PZT to vertically vibrate the resonator. The probe and pump transmission light from the WGM resonator were separated using a wavelength division multiplexer (WDM), and their output powers were individually sensed by an avalanche photodiode (APD). The DC and AC components of the probe output were separated by a bias-tee, and they were detected by an oscilloscope and spectrum analyzer, respectively. The probe frequency was scanned only when the transmission spectra is obtained and was locked to the HWHM of the probe resonance during the optomechanical measurements. The DC output of the pump laser was monitored by an oscilloscope to figure out the intracavity pump power.

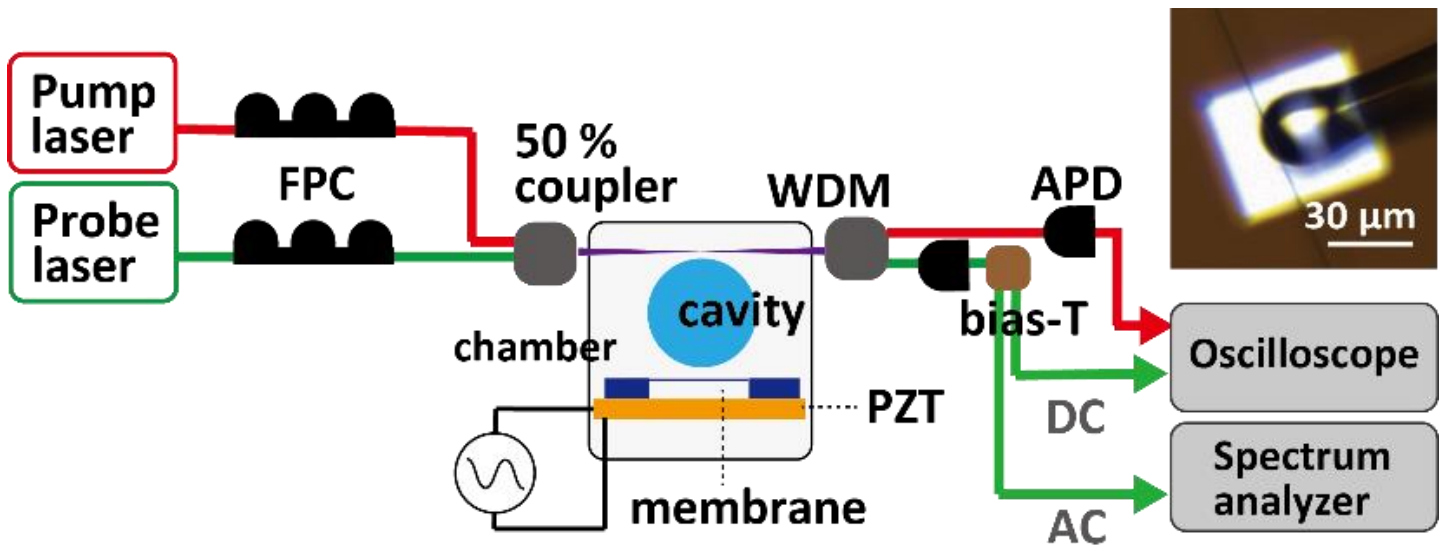

Fig. 3. Experimental setup. Probe and pump lasers are evanescently coupled to the cavity via a tapered fiber. The transmitted laser is separated using a WDM and sensed with an APD. The probe transmission light is then separated into DC and AC components and detected with an oscilloscope and spectral analyzer. The figure inset is a microscopic image of the fiber, cavity, and resonator.

## 3. Results

### *3.1 Probe transmission with CW-CCW hybridized optical mode*

To clarify the improvement of the optical Q factor and changes in the transmission spectrum of the probe resonance with Raman gain, we measured the probe transmission spectra with increasing the pump power. Fig. 4(a) represents experimentally obtained probe transmission spectra initiated in the under-coupling state without Raman gain. The intracavity pump power coupled to the WGM resonator was increased from 0.01 to 0.97 mW by changing the pump detuning. Note that a high input power generates thermal broadening of the resonance mode and induces a sawtooth-like spectrum in scanning from shorter to longer wavelength [12,25]. The intracavity pump power is thermally locked and stable at the arbitrary detuning owing to the broadened pump resonance spectrum. The intracavity pump power was estimated from the output DC power measured by an oscilloscope.

The transmission spectrum obtained in the experiment showed mode splitting that is generated owing to the ultrahigh-Q resonance where the small refractive index inhomogeneity of the cavity surface resolves the degeneracy between clockwise and counterclockwise (CW-CCW) optical modes. The light transmission of the splitting WGM can be described using the input-output theory [14]:

$$T \approx \left|1 - \frac{\kappa_{\mathrm{ex}}\beta}{\beta^2 + g^2}\right|^2, \tag{1}$$

with $\beta \equiv -i(\Delta\omega - g) + (\kappa_{\mathrm{eff}} + \kappa_{\mathrm{ex}})/2$, where $g$ and $\Delta\omega$ denote the coupling strength between CW-CCW optical modes and probe detuning ($\Delta\omega = \omega_{probe} - \omega_0$). At the coupling state without Raman gain, the intrinsic loss, coupling loss, and CW-CCW coupling strength were 15, 11, and 17.5 MHz, respectively, calculated by curve fitting using Eq. (1). With increasing pump power, linewidth narrowing was observed as well as the shift of coupling from the under- to critical-coupling state. A 1.5 times higher optical Q factor was achieved with 0.97-mW intracavity pump power via Raman gain [see blue solid line in Fig. 4(a)]. Fig. 4(b) shows the calculated probe transmission spectra with different Raman gain rates $\xi$ at the under-coupling state. The initial waveguide-cavity coupling condition was set to be the same as the measurement result. The effective loss can be obtained by curve fitting of the transmission spectrum. The coupling loss and CW-CCW coupling strength remain the same because the cavity-waveguide coupling condition does not change with the pump input. The calculated transmission spectra of the CW-CCW hybridized optical mode were in a good agreement with the experimental results. Based on the calculated transmission of the CW-CCW hybridized optical mode, the Raman amplification rate of $\xi = 1.0\kappa_0$ induces the minimum transmission, while $\xi = 0.26\kappa_0$ gives the critical coupling state for the single optical mode. The Raman amplification rate $\xi$ that gives the critical-coupling state is fourfold larger than that for the single optical mode because of the additional optical energy transfer between the CW and CCW modes.

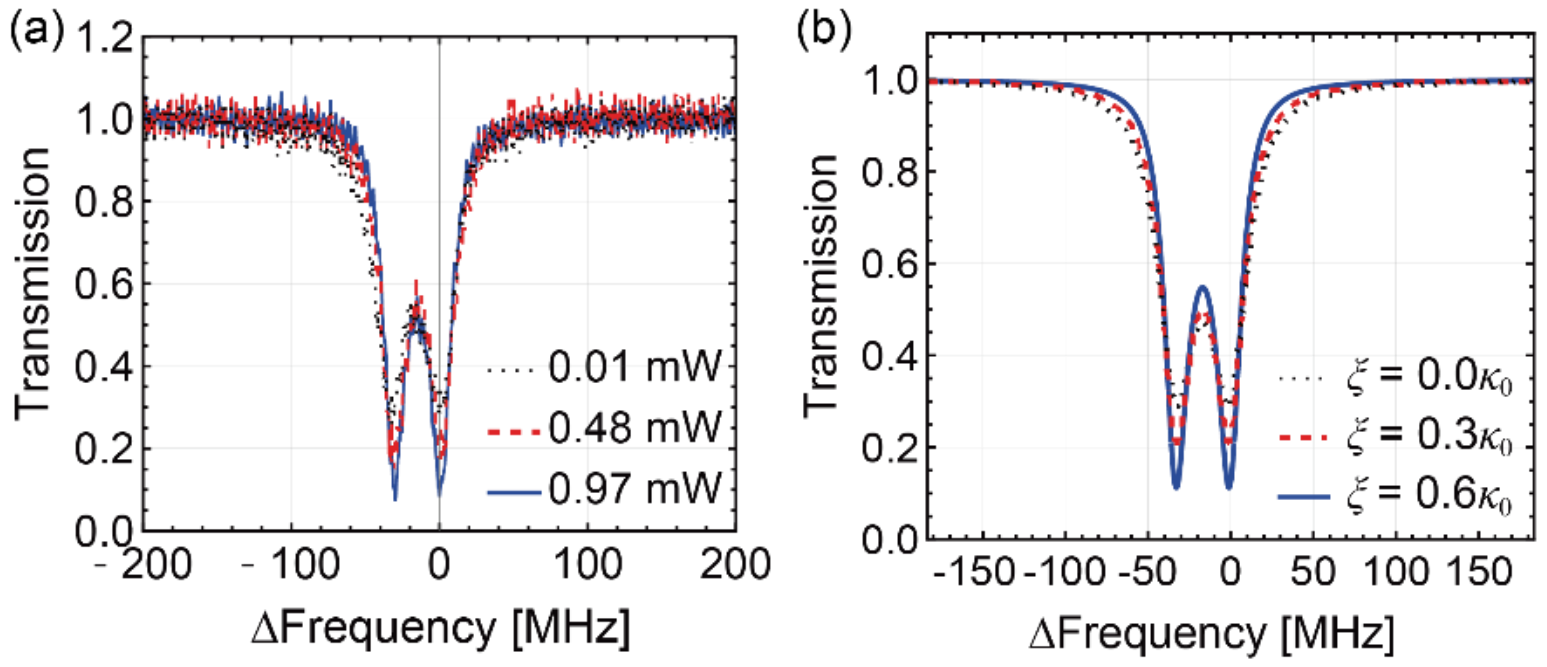

Fig. 4 (a) Experimentally obtained transmission spectra initiated in the under-coupling state. The intracavity pump powers coupled to the WGM resonator were 0.01 (black dotted line), 0.48 (red dashed line), and 0.97 mW (blue solid line). (b) Changes in the calculated transmission spectra of the probe laser initiated in the under-coupling state with $\xi$ = 0, 0.3, and 0.6 $\kappa_0$. Intrinsic $\kappa_0$ and coupling losses $\kappa_{ex}$ used in the calculation were 15 and 11 MHz, respectively.

### *3.2 Raman gain-enhanced near-field optomechanical transduction*

To characterize basic properties of near-field optomechanical coupling between the optical cavity and membrane resonator, we measured the optical transmission spectra with respect to the cavity-membrane gap without Raman gain. Probe transmission spectra at ~1570 nm for different cavity-membrane gaps are shown in Fig. 5(a). The optical quality factor of Q = $4.2\times10^7$ was almost unchanged, and the optomechanical coupling rate, given by $G(x_0) = \mathrm{d}\omega_0(x)/\mathrm{d}x|_{x=x_0}$[26], increased exponentially as the cavity-membrane gap decreased, as shown in Fig. 5(b). The optomechanical coupling between the cavity and resonator was purely dispersive, in which the optical response is described by a frequency shift caused by the change in the effective refractive index.

The power spectral density (PSD) of the mechanical vibration of the resonator detected by the spectrum analyzer is associated with changes in the laser detuning for the probe resonance; therefore, the maximum power spectral density can be obtained with the probe detuning near the HWHM of the probe resonance, as described in subsection 2.1. Fig. 5(c) shows the PSD of the mechanical vibrations at different probe detuning frequencies. The mechanical resonance frequency was 2.265 MHz, and the mechanical Q factor was $9.1\times10^1$ under ambient pressure [see the black line in Fig. 5(d)]; it was lower than that in vacuum owing to the viscous damping. The PSD was at its maximum with Δ ~7.8 MHz—which was almost at the HWHM of the probe-resonance spectrum. The change in PSD against the detuned frequency was asymmetrical owing to the CW-CCW hybridized optical mode.

Finally, we measured the mechanical vibration spectra of the membrane resonator with Raman gain to characterize the change in PSD. The probe detuning was locked at $\Delta \sim 7.8$ MHz, and the pump frequency was detuned by 0–75 GHz to change the pump-input power. Fig. 5(d) represents the PSD of the mechanical vibrations of the membrane resonator for a frequency range of 2.20–2.33 MHz with (red) and without (black) Raman gain. It is obvious that the Raman gain improves PSD in the optomechanical transduction. This improvement in PSD purely comes from the Raman gain; it does not originate from the optomechanical backaction, because there is no frequency shift and linewidth change in either spectrum. The black-square plot in Fig. 5(e) shows the change in PSD at 2.265 MHz with various intracavity pump powers after subtracting the frequency-independent noise floor. It was normalized with respect to that without Raman gain. By increasing the intracavity pump power from 0.30 to 1.25 mW, the PSD was increased fourfold. The intracavity pump power of 1.25 mW is a technical limit because higher pump power causes a mode hop of the pump light owing to the extremely broadened pump-resonance mode by the photothermal effect.

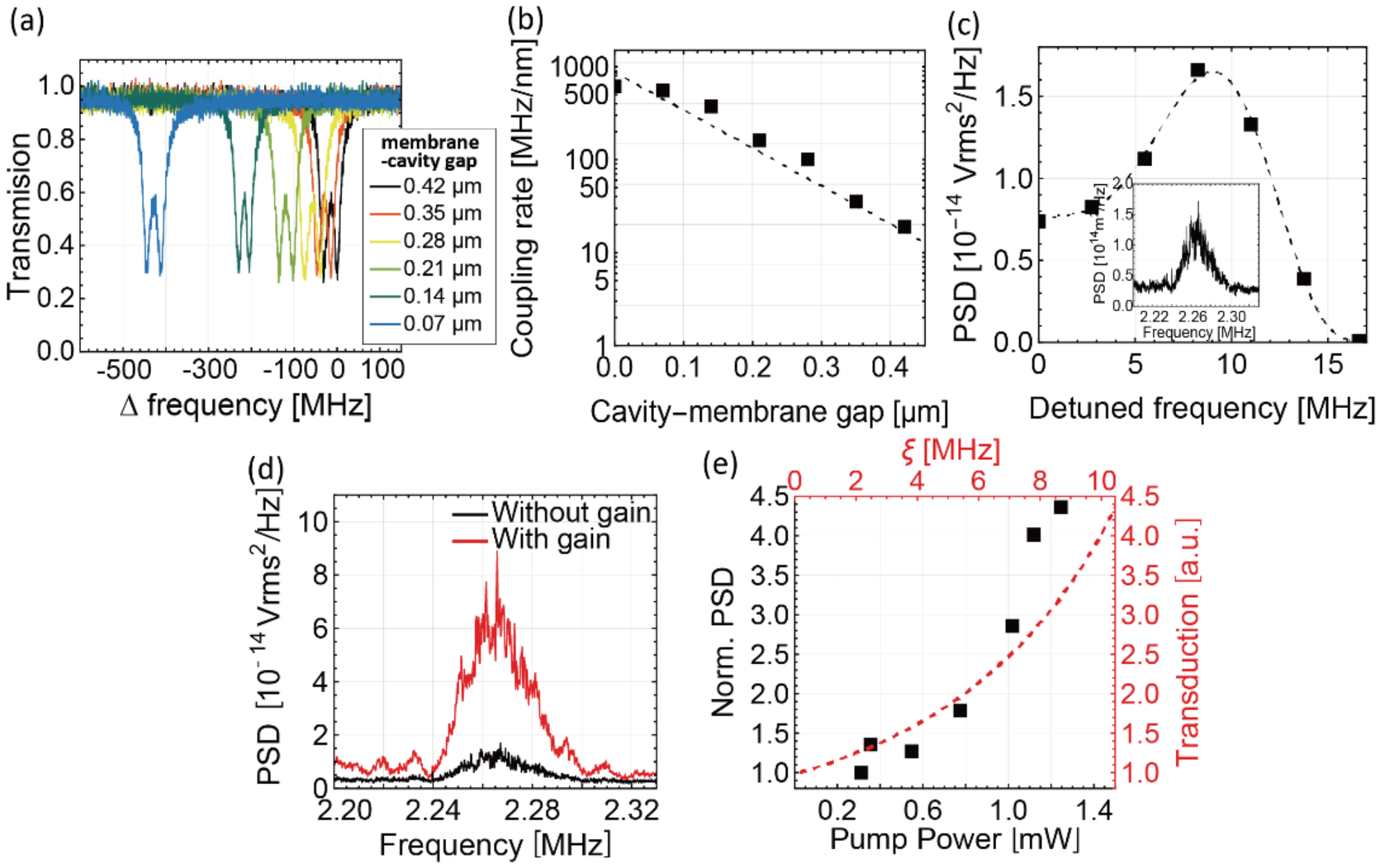

Fig. 5. (a) Changes in probe transmission spectra for cavity-membrane gaps of 0.07–0.42 μm. (b) Optomechanical coupling rate calculated from the frequency shift as a function of cavity-membrane gap distance. (c) PSD of the mechanical vibration spectrum for probe detuning of $\Delta$ = 0–17 MHz. (d) PSD of mechanical vibrations without (black) and with (red) Raman gain. The intracavity pump power was 1.25 mW for the Raman amplification process. A random-noise input of 200 mV was applied to the PZT attached to the membrane resonator. (e) PSD above the noise floor of mechanical-vibration spectra for an intracavity pump power of 0.30–1.25 mW (black-square plot) and calculated transduction coefficient $\eta$ (red-dashed line). The intrinsic and coupling losses used for the calculations were 15 and 11 MHz, respectively. The $\xi$ (upper x-axis) was calculated from the probe transmission spectra, and the transduction coefficient $\eta$ was normalized with respect to that at $\xi = 0$ Hz.

## 4. Discussion

To clarify the dependence of the transduction coefficient on the Raman gain beyond the experimental restriction, we theoretically estimated $\eta = P_{\mathrm{cav}}^2 Q^2$. The red-dashed line in Fig. 5(e) represents the theoretically estimated transduction coefficient of the CW-CCW hybridized optical mode depending on the Raman amplification rate $\xi$ (upper x-axis). The initial coupling condition was set to be the under-coupling with $\kappa_0$ and $\kappa_{ex}$ of 15 and 11 MHz, respectively, which are the same as the experimentally obtained loss rates. The transduction coefficient $\eta$ increases with the Raman amplification rate $\xi$ in a range of $\xi < \kappa_0$. The experimental results [black-square plots in Fig. 5(e)] demonstrate that the change in the PSD was qualitatively consistent with the calculated transduction coefficient in the experimentally achievable regime ($0 < \xi < 8$ MHz).

To quantitatively explain the advantage of using Raman gain in optomechanical coupling detection, we performed theoretical calculations of changes in optomechanical transduction coefficient in different initial coupling states and at different Raman amplification rates. The changes in the transduction coefficient with single and CW-CCW hybridized optical modes are shown in Fig. 6(a) and (b), respectively. The coefficient was normalized with respect to $\eta$ at the condition of $\kappa_{ex} = \kappa_0$ and $\xi = 0$, that is, $\eta_c$. The dashed line represents the states at $\eta = \eta_c$; the regions colored red represents the effectively amplifiable conditions. It indicates that the change in the transduction coefficient depends largely on the initial coupling states. The optical Q increases with $\xi$ regardless of the initial coupling states; however, the intracavity probe power also determines the optomechanical transduction.

For the single optical mode in a range of $0.2 < \kappa_{ex}/\kappa_0 < 1.0$ [see Fig. 6(a)], the transduction coefficient $\eta$ is larger than $\eta_c$ and increases with $\xi$ owing to the increase in the intracavity probe power. On the contrary, smaller $\eta$ was obtained with $\kappa_{ex}/\kappa_0 > 1.0$ because of the decreased intracavity probe power at the over-coupling state. Generally, $\eta$ in the over-coupling state is more than an order of magnitude smaller than that in the under-coupling state. The intracavity probe power starts decreasing with increasing $\xi$ at $\kappa_{ex}/(\kappa_0 - \xi) = 1$ and gradually offsets the increase in optical Q. The CW-CCW hybridized optical mode exhibits similar transduction characteristics. The largest difference is the delayed transit from effective amplifiable (red) to negative amplifiable (blue) conditions as $\xi$ increases. The effective amplifiable condition can be accessed even at the over-coupling state ($\kappa_{ex}/\kappa_0 \lesssim 2.2$). For both single and CW-CCW hybridized optical modes, our setup would demonstrate that large improvement can be obtained by executing the maximum amplification below the onset of self-sustained oscillation to an initially under-coupled state (right-bottom region).

We next discuss the role of shot noise induced by the amplification process. Displacement measurement in a standard near-field optomechanical setup [17-19] is dominated by optical shot noise, i.e., photon number deviation in the probe laser. The Raman amplification process generates additional photon number deviation by the spontaneous Raman scattering. It cannot improve the shot-noise-limited sensitivity owing to the fluctuation-dissipation theorem [27]. In fact, in our experimental results, we observed an increased noise floor level of the vibration spectrum measured with a larger pump power [see Fig. 5(d)]. The most intrinsic contribution was the spontaneous Raman scattering. With a higher Raman amplification rate, the change in the noise floor level owing to the noise induced by the pump laser would dominate the frequency noise and alter the characteristics of the optomechanical transduction.

Although the optomechanical signal transduction in terms of the photon number cannot be improved owing to the additional noise, that in terms of the optical cavity Q can be improved by the Raman gain, as discussed above. Consequently, the signal-to-noise ratio can be totally increased at least in the shot-noise-limited regime. In fact, the measured signal-to-noise ratio in Fig. 5(d) is still improved by a factor of 2.2, showing that the Raman amplification is effective even in the shot-noise-limited regime. The other major technical contribution to the noise was a periodic fluctuation (~10 kHz) of the cavity itself as a cantilever [26,27]. It can be removed by further improving experimental setup to include feedback cooling scheme [28,29].

Finally, we emphasize that our optical-gain-assisted optomechanical setup is applicable for not only displacement measurement but also mechanical control of active optical states. The near-field optomechanical platform allows us to readily merge active optical nonlinear processes, including lasing with mechanical nonlinear processes such as bifurcation and chaos [30,31]. This highly tailorable optomechanical platform will pave the way to the field of active optomechanics with various materials and structures.

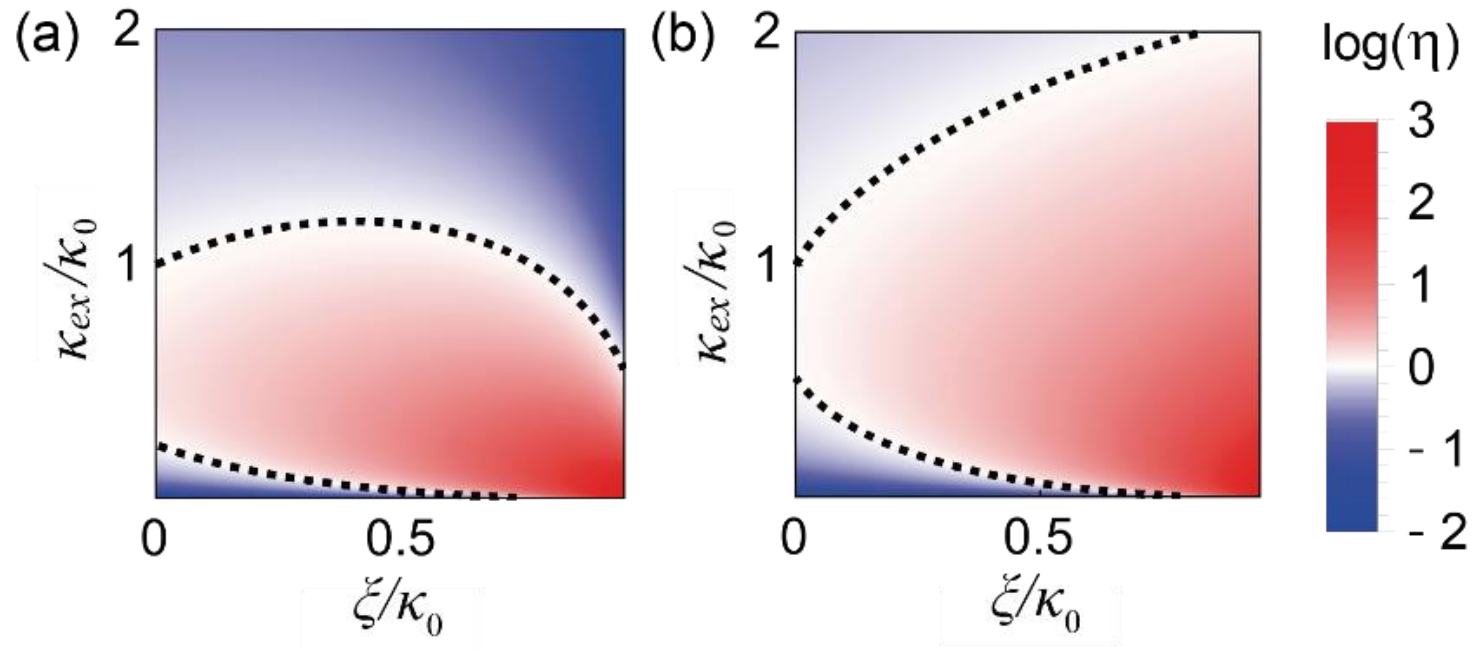

Fig. 6. Transduction coefficient for $\xi/\kappa_0$= 0–0.95 and $\kappa_{ex}/\kappa_0$ = 0.1–2.0 of (a) single and (b) CW-CCW hybridized optical modes. The $\eta$ was normalized with respect to that at the critical coupling without Raman gain $\eta_c$. The dashed line shows the state at $\eta = \eta_c$.

## 5. Conclusion

We have demonstrated Raman-gain-enhanced optomechanical coupling between a thin SiN-membrane resonator and movable silica WGM resonator. The Raman amplification can control the optical Q and intracavity probe power. It results in the improved optomechanical transduction coefficient, especially in the under-coupling state. In our work, the PSD of the mechanical-vibration spectrum was quadrupled and in good agreement with the calculated transduction coefficient. The transduction coefficient can be further improved by using an optical cavity with a smaller mode volume. It could also be improved by introducing optical gains by using the rare earth ions, including erbium- or ytterbium-doped WGM resonator [25,32,33], in which a higher optical gain can be achieved with a lower pump-input power. The movable WGM resonator with Raman gain-enhanced transduction allows a variety of ultrasensitive displacement measurements with free-standing optical resonators. The extension of this near-field optomechanical coupling technique will give rise to new and effective sensing and control methods for optical, mechanical, and electrical oscillators.

**Acknowledgments.** This work was supported by JSPS KAKENHI (21H01023, 23H05463).

**Disclosures.** The authors declare no conflicts of interest.

**Data availability.** Data are available from the corresponding author upon reasonable request.